\documentclass[prl,showpacs,twocolumn,aps]{revtex4}

\usepackage{amsmath}
\usepackage{amssymb}
\usepackage{latexsym}
\usepackage{amsfonts}
\usepackage{epsfig}
\usepackage{psfrag}
\usepackage{graphicx}

\usepackage{color}

\newcommand{\nn}{\nonumber\\}

\newcommand{\f}[1]{\mbox{\boldmath$#1$}}

\newcommand{\vau}{\mbox{\boldmath$v$}}
\newcommand{\na}{\mbox{\boldmath$\nabla$}}
\newcommand{\bea}{\begin{eqnarray}}
\newcommand{\ea}{\end{eqnarray}}
\newcommand{\eea}{\end{eqnarray}}

\newcommand\Om{\Omega}
\newcommand\vp{\varphi}

\def\d{\partial}

\newcommand\ra{\rightarrow}
\newcommand\w{\omega}
\def\ra{\rightarrow}

\begin{document}

\title{Whispering gallery like modes along pinned vortices}

\author{Piotr Marecki and Ralf Sch\"utzhold}

\affiliation{Fakult\"at f\"ur Physik, Universit\"at Duisburg-Essen,
Lotharstrasse 1, 47057 Duisburg, Germany}

\date{\today}

\begin{abstract}
Employing Unruh's analogy to gravity, we study sound propagation in
stationary and locally irrotational vortex flows where the circulation
is wound around a long (rotating) cylinder.
Apart from the usual scattering solutions, we find anomalous modes
which are bound to the vicinity of the cylinder and propagate along
its axis -- similar to whispering gallery modes.
For supersonic flow velocities (corresponding to an effective ergoregion),
these modes can even have zero frequency.
Thus they should be relevant for the question of stability or
instability of this set-up.
\end{abstract}


\maketitle

\paragraph*{Introduction}

The full characterization of sound modes propagating within a given flow 
profile is a major problem in fluid dynamics and often reveals 
very rich physics.
Even for stationary flows, which admit a separation ansatz where the 
linear perturbations can be labeled by their conserved frequency $\w$, 
this problem is highly non-trivial:
For \emph{static} systems, depending on what sort of 
 scenario is considered, 
the dynamics of perturbations is governed by equations  
of Sch\"odinger 
$i\partial_t\psi={\cal H}\psi$ 
or d'Alembert 
$\partial_t^2\Phi={\cal D}\Phi$ 
type. 
In such cases the full characterization of solutions follows from the 
spectral analysis of the differential operators ${\cal H}$ or ${\cal D}$.
For \emph{stationary} systems, however, the following inherent difficulty 
appears.
The equations assume the form 
$\partial_t^2\Phi+{\cal A}\,\partial_t\Phi={\cal B}\Phi$,
where the two  operators ${\cal A}$ and ${\cal B}$
do not commute in general, and therefore their spectral content 
has no direct significance for the problem at hand. 
As a result, questions like the completeness of solutions or the
existence of unstable modes with $\Im(\omega)<0$ are far more 
difficult to address. 

The wave equation for sound in a locally irrotational and 
stationary background flow has the form mentioned above,   
$\partial_t^2\Phi+{\cal A}\,\partial_t\Phi={\cal B}\Phi$.
Precisely the same structure arises for scalar fields propagating 
in a stationary space-time. 
Moreover, as discovered by Unruh \cite{Unruh}, there is an exact
analogy between the two cases: 
Let us consider a fluid with density $\varrho$ and velocity $\vau$, 
whose pressure $p$ is a function of $\varrho$ only $p=p(\varrho)$, 
i.e., the fluid is barotropic. 
The perturbations, i.e., sound waves, can be parametrized by 
a single potential $\Phi$ via $\delta\vau=\na\Phi$ and 
$\delta\varrho=\varrho\dot\Phi/c^2_{\rm s}$.
Neglecting viscosity, they obey the same wave equation as a scalar 
field in a curved space-time described by the effective acoustic 
metric \cite{Unruh}
\bea
\label{metric}
ds^2=\frac{\varrho}{c_{\rm s}}
\left(c^2_{\rm s}dT^2-[d\f{R}-\vau\,dT]^2\right)
\,,
\ea
where $T$, $\f{R}$ are the laboratory coordinates and $c_{\rm s}$
is the speed of sound $c_{\rm s}^2=dp/d\varrho$.
This analogy allows us make use of many geometrical tools and concepts
of general relativity in fluid dynamics.

\paragraph*{Vortex flow}

In the following, we consider a stationary and locally irrotational 
$\na\times\vau=0$ flow around a long cylinder.
Aligning the coordinate $Z$-axis with the symmetry axis of the cylinder,
we assume the flow velocity to be 
$\vau=v(R)\f{e}_\varphi$ in cylindrical coordinates $Z,R,\varphi$.
The condition $\na\times\vau=0$ then implies $\vau=\f{e}_\varphi\kappa/R$,
where $\kappa$ determines the circulation.

For normal fluids, such a profile approximates the stationary flow
around a rotating cylinder.
Indeed, if the fluid is incompressible 
$\na\cdot\vau=0$, 
the above velocity profile provides an exact solution of the Navier-Stokes 
equations (similar to a tornado away from the core).
While this solution gets modified in more realistic models of normal fluids, 
for superfluids (such as $^4$He II at low temperatures), vorticity can only 
occur in the form of vortices with a quantized circulation, which is thus 
topologically stabilized and not so easy to create (or destroy).
The flow profile $\vau=\f{e}_\varphi\kappa/R$ then corresponds to a
vortex which is pinned around a long cylinder (e.g., wire), where
$\kappa$ is an integer multiple of 
$\kappa_0=\hbar/M_{\rm He}$ \cite{Donnelly_book}.

In fluid dynamics, it is often useful to express the problem in terms
of  dimensionless quantities (such as Rossby number) in order to
exploit the scaling symmetries.
Here, we do the same and choose the sound velocity at infinity
$c_{\rm s}(R\uparrow\infty)\equiv c_\infty$ as reference scale.
Using the circulation $\kappa$ we get a length
${\mathfrak L}=\kappa/c_\infty$ and a time scale
${\mathfrak T}=\kappa/c_\infty^2$.
By rescaling the laboratory coordinates $T$ and $\f{R}$, to get
dimensionless $t$ and $\f{r}$, we also find that
(\ref{metric}) becomes
\bea
\label{line-element}
ds^2
=
\frac{\rho}{c}
\left[
\left(c^2-\frac{1}{r^2}\right)dt^2
+2dt\,d\varphi
-dr^2-r^2d\varphi^2-dz^2
\right]
\ea
where $c(r)=c_{\rm s}(r)/c_\infty$ is the dimensionless sound
speed which approaches unity at infinity.
Similarly, the normalized density reads
$\rho(r)=\varrho(r)/\varrho_\infty$.
Typically, both decrease a bit when approaching the cylinder
where the velocity $\vau$ increases and thus the pressure
$p$ drops.

The coordinates $t,z$ and $\varphi$
have their standard ranges, but $r$ is restricted to
$r\in(r_{\rm w},+\infty)$, where $r_{\rm w}$ is the re-scaled
wire radius.
Note that the acoustic metric (\ref{line-element}) possesses an
{\em ergoregion} $g_{00}<0$ for small enough radii
$r<1/c$, if such are in the allowed range, i.e., if
$r_{\rm w}<1/c(r_{\rm w})$.
Appearance of ergoregion means that the flow velocity $\vau$ 
exceeds the local speed of sound $c_{\rm s}$ somewhere 
(e.g., near the wire).
As we shall see, the presence an ergoregion has
profound consequences for the sound modes.

\paragraph*{Geometric acoustics}

According to Unruh's analogy to gravity \cite{Unruh}, sound modes in the
vortex profile are solutions of the wave equation of a massless scalar
field in the metric (\ref{line-element}).
However, before investigating the full wave equation, let us get some
insight via the WKB approximation -- which amounts to studying
sound rays.
They are null geodesics in the space-time (\ref{line-element})
which can be found in complete analogy to the general relativistic
Kepler problem.
We search for sound rays $x^a(\tau)$,
and find that the problem is reduced to quadratures due
to existence of four independent first integrals.
The space-time (\ref{line-element}) admits three  symmetries
with the Killing vectors $\partial_t$, $\partial_\varphi$,
and $\partial_z$.
Via the Noether theorem, this implies the conservation of
the energy $E=\Om(f\dot t + \dot \vp)$,
the angular momentum $J=\Om(r^2\dot \vp - \dot t)$,
and the axial momentum $P=\Om \dot z$,
where the dot denotes $d/d\tau$.
Here the notation $\Om=\rho(r)/c(r)$ and
$f=c^2-1/r^2$ is introduced for brevity.
Together with the null ray condition $\dot x_a\dot x^a=0$,
we can express all velocities in terms of these first integrals,
e.g., $\dot t=(E-J/r^2)/(\Om c^2)$ and $\dot\vp=(E+fJ)/(\Om c^2 r^2)$.
The remaining radial equation reads
\bea
\label{radial}
\Om^2\dot r^2+P^2
=
\frac{E^2}{c^2} - \frac{J^2c^2+2EJ}{c^2r^2} + \frac{J^2}{c^2r^4}
=
E^2 - V_{\rm eff}(r)
\,,
\ea
where we have introduced the effective potential $V_{\rm eff}(r)$
which also contains the term $E^2(1-1/c^2)$.

The sound rays can now be classified by the
following arguments.
For $r\uparrow\infty$, the effective potential $V_{\rm eff}(r)$
vanishes.
Hence all scattering solutions must have $E^2\geq P^2$.
For $r\downarrow0$, of the other hand, the effective potential
$V_{\rm eff}(r)$ diverges $V_{\rm eff}(r\downarrow0)\downarrow-\infty$
for $J\neq0$.
Thus rays are strongly attracted by the vortex in its vicinity. 
There is a cut-off in $r$, however, provided by $r_{\rm w}$ 
(wire radius) where the sound rays are reflected.
If, due to $r_{\rm w}$, the potential $V_{\rm eff}(r)$ is
monotonically \emph{decreasing} for all $r>r_{\rm w}$, then only 
scattering solutions ($E^2\geq P^2$) exist.
If, however, there are local minima of $V_{\rm eff}(r)$ at finite 
$r\in[r_{\rm w}, \infty)$, bound rays oscillating around them will exist.
As one may easily infer from the structure of Eq.~(\ref{radial}), 
this can always be achieved by tuning the angular momentum $J$.
Choosing, e.g., $J=-E$, we see that $V_{\rm eff}(r)$
is strictly negative (assuming $c\leq1$ everywhere; see below), 
and that there will exist rays bouncing off $r_{\rm w}$ indefinitely.
Note that these are counter-rotating rays, i.e.,
propagating against the vortex flow.

In the special case of constant $c$ and $\rho$ (i.e., $\Om=1$),
these qualitative arguments can be made precise.
The maximum of $V_{\rm eff}(r)$ is at
$r_*=\sqrt{2J/(J+2E)}$ where $V_{\rm eff}'(r_*)=0$.
Let $J$, $E$ be fixed and $r_{\rm w},P$ adjustable. 
For $r_{\rm w}>r_*$ we only have scattering rays, 
otherwise there exist also bound rays.
Generally, for arbitrarily large wire radii $r_{\rm w}$, 
one can find values of $J$ and $E$ for which $r_*$ is imaginary, 
and therefore bound rays exist.
For small $J$, the radius $r_*$ goes to zero -- i.e.,
bound states require a minimum angular momentum.
Finally, for large $J$, the radius $r_*\to\sqrt{2}$
which is outside the ergoregion at $r=1$.

Surprisingly, in the special case of constant $c$ and $\rho$
(i.e., $\Om=1$) as above, the problem of finding null orbits, $r(\vp)$,
is exactly soluble in terms of elliptic functions (see \cite{PM_full}),
the reason being not-more-than quartic dependence of $V_{\rm eff}(r)$
on $1/r$, as is also the case in the general relativistic Kepler
problem, for example.
We note that the $J^2/r^4$-term is crucial for the universal
existence of bound states as discussed above.
As we shall see later, this remains correct for the full wave equation.
However, available treatments 
\cite{Fetter_sound_scattering,Stone_Iordanskii,Fischer_Flaig}
of the subject of propagation of waves in vortex backgrounds introduce 
assumptions, which effectively eliminate this term. 
While this leads to the radial equation of simple Bessel form, 
it also is the reason why the family of bound states reported upon here 
is not to be found in the literature
\cite{Fetter_sound_scattering,Stone_Iordanskii,Fischer_Flaig}. 

\paragraph*{Wave acoustics}

In what follows the full wave equation will be considered.
To this end, let us first discuss the field expressions for $E$, $J$,
and $P$.
They can be obtained from the pseudo energy-momentum tensor
\cite{pseudo}
\begin{equation}
T_{ab}[\Phi]=(\d_{a}\Phi)(\d_{b}\Phi)-\frac{1}{2}\,g_{ab}\,g^{cd}\,
(\d_c\Phi)(\d_d\Phi)
\,,
\end{equation}
where $\Phi$ is the velocity potential $\delta\vau=\na\Phi$
of the sound waves $\delta\vau$ and $g_{ab}$ is the acoustic metric
(\ref{line-element}).
For each Killing vector field $\xi$, we get a conserved field quantity
$\Xi=\int dS^a\,T_{ab}\,\xi^b$
via integrating over the spatial hyper-surface $dS^a$.
As usual, invariance under time-translations leads to the conserved energy,
\bea
E[\Phi]
=
\int d^3r\,\frac{\rho}{2 c^2}
\left(\dot\Phi^2+c^2 [\na\Phi]^2-\frac{1}{r^4}[\d_\vp\Phi]^2\right)
\,.
\ea
where Minkowski products are implied in the term $[\na\Phi]^2$, i.e.,
$[\na\Phi]^2=(\d_r\Phi)^2+(\d_z\Phi)^2+(\d_\vp\Phi )^2/r^2$.
The energy functional is positive definite as long as
\mbox{$c^2_{\rm s}>\vau^2$} everywhere, i.e., $c^2>1/r_{\rm w}^2$.
This is not anymore the case if the ergoregion belongs to
the spacetime.
The two remaining conserved functionals following from the
Killing vector fields $\partial_z$ and $\partial_\varphi$,
are the axial momentum
\bea
P[\Phi]
=
\int d^3r\,\frac{\rho}{2 c^2}
\left(\d_t\Phi+\frac{1}{r^2}\d_\vp\Phi \, \right)\, \d_z\Phi
\,,
\ea
and similarly the angular momentum $J[\Phi]$ with $\d_z\Phi\to\d_\vp\Phi$.
Furthermore, due to the $U(1)$ gauge-invariance of the complexified
wave equation, the Klein-Fock-Gordon inner product of two solutions
\bea
(\Phi_1|\Phi_2)
=
\frac{i}{2}
\int d^3r\, \Phi_1^*
\overleftrightarrow{\left(\d_t+\frac{1}{r^2}\d_\vp\right)}
\Phi_2
\,,
\ea
with $\Phi_1^* \overleftrightarrow{ \d_a} \Phi_2=
\Phi_1^*\d_a\Phi_2-\Phi_2\d_a\Phi_1^*$
is conserved, i.e., independent of the Cauchy surface over which it is taken.

\paragraph*{Separation ansatz}

In view of the symmetries of our set-up, we approach the problem of solving
the wave equation by considering modes specified by the following ansatz,
which reflects the structure of the Killing vectors
\bea
\label{separation}
\Phi(t,r,\varphi,z)=\phi(r)\exp\{-i\omega t+im\varphi+ip_zz\}
\,.
\ea
For such modes the conserved quantities are related to the inner product via
$E[\Phi]=\omega(\Phi|\Phi)$,
$P[\Phi]=p_z (\Phi|\Phi)$ and
$J[\Phi]=m (\Phi|\Phi)$.
Since $(\Phi|\Phi)$ and $E[\Phi]$ are always real, solutions with complex
frequencies must have $(\Phi|\Phi)=0$ and $E[\Phi]=0$.
In the absence of an ergo\-region, $E[\Phi]$ is positive definite
and all frequencies are real, i.e., the flow is linearly stable
\footnote{This argument can be made more precise.
For $r_{\rm w}>1$, a Hamiltonian formulation of the problem exists
\cite{PM_full} with the Hamiltonian $H$ being a self-adjoint operator
acting on a Hilbert space.
Thus in this case all $\w$ (eigenvalues of $H$) are real, and the
family of corresponding modes is complete in the usual sense.
For $r_{\rm w}<1$, on the other hand, the operator $H$ is only a
symmetric operator on a Krein space, and the presence of complex
$\w$ cannot be excluded {\em a priori}.
However, based on the $(p_z,m,r_{\rm w})$-dependence of real $\w$,
we conjecture that they do not appear \cite{PM_full}.
To the best of our knowledge, neither the problem of existence of
complex frequencies, nor the issue of completeness of the eigenmodes
of $H$ has been settled in the literature for this case.}.
Furthermore, the frequency $\omega$ and the pseudo-norm $(\Phi|\Phi)$
of our modes have the same sign in this case.
As a result, creation and annihilation operators are associated to
modes with positive and negative frequencies, respectively.
In the case with an ergoregion, the energy can become negative and
hence this  property is no longer true.
This can lead to interesting and related phenomena such as super-radiance
\cite{Unruh_superradiant} and the Klein paradox \cite{Fulling_book}.
Since a given frequency $\omega>0$ can be associated
to both, creation and annihilation operators, one can have a mixing
of the two and thus phenomena like particle creation.

By inserting the above separation ansatz (\ref{separation}) 
into the wave equation we reduce it to a single ordinary differential
equation (in radial direction):
%
%
\bea
\label{eigenvalue}
\left[
-\frac{1}{r\rho}\,\frac{d}{dr}\,
r\rho\,\frac{d}{dr}
+\omega^2\left[1-\frac{1}{c^2}\right]
+\frac{m^2c^2+2m\omega}{c^2r^2}
-\frac{m^2}{c^2r^4}
\right]\phi
\nn
=
{\cal H}\phi
=
\left(
{\omega^2}-p_z^2
\right)\phi
=
\lambda\phi
\,.
\quad
\ea
%
%
This main equation of our sound-propagation problem has several
interesting features \cite{PM_full}.
First of all, because $\rho\ra 1$ and $c\ra 1$ at $r\uparrow\infty$,
the solutions $\phi(r)$ at large $r$ are either oscillating,
for $\w^2>p_z^2$, or exponentially decaying, for $\w^2<p_z^2$.
In complete analogy to the ray problem,
we call the first type of solutions the \emph{scattering modes},
and the second the \emph{bound-state modes}.
Finding these modes is then reduced to an eigenvalue problem
${\cal H}\phi=\lambda\phi$ with ${\cal H}={\cal D}+{\cal V}$,
where ${\cal D}$ stands for the ``kinetic part''
involving $r$-derivatives and ${\cal V}$ is the effective
potential.
Note that $V_{\rm eff}(r)$ in the sound-ray problem corresponds to ${\cal V}$ on identification of $E$ and $J$ with $\omega$ and $m$, repectively.
For the scalar product 
\bea
\label{scalar}
\{\phi_1|\phi_2\}=\int\limits_{r_{\rm w}}^\infty dr\;\rho(r)r\,
\phi_1^*(r)\phi_2(r)
\,,
\ea
the operator $\mathcal H$ is self-adjoint if the Neumann
boundary condition at $r_{\rm w}$ is assumed $\phi'(r_{\rm w})=0$,
which would just reflect the fact that perturbations can not penetrate the wire.
Even though neither is $\cal H$ the Hamiltonian of the problem,
nor is the scalar product (\ref{scalar}) distinguished by the
geometry, both of these elements suffice for the following general statements.
First of all, ${\cal H}$ has bound states, $\lambda<0$, only if it one can
find test functions $\psi$ such that $\{\psi|{\cal H}|\psi\}<0$.
The kinetic part, ${\cal D}$, is always non-negative, and therefore bound
states can only exist if ${\cal V}$ is sufficiently negative.
A lower bound on the eigenvalue $\lambda$ can be obtained via
the minimum of ${\cal V}(r)$.
Then, for general profiles of $c(r)$ and $\rho(r)$ with the
aforementioned asymptotics, we get the following statements:
\\
{\bf 1)}
Bound states with $\omega=0$ can only exist if the velocity becomes
supersonic somewhere $c<1/r$, i.e., if the acoustic spacetime has an
ergoregion, for otherwise ${\cal V}(\omega=0)$ is non-negative --
consistent with $E[\Phi]=\omega(\Phi|\Phi)$.
\\
{\bf 2)}
Bound states with $m=0$ can only exist if $c_s$ becomes
sufficiently smaller than $c_\infty$ near the wire.
The mechanism for bound states in this case is just the total
reflection form the region with larger speed of sound --
which can also occur in a non-rotating fluid.
\\
{\bf 3)}
Independently of the mechanism of total reflection,
caused by the $\w^2$ term in $\cal V$, the other terms in $\cal V$
allow for bound states with $m\omega>0$ (i.e., co-rotating)
only  in the presence of an ergoregion,
i.e., if the flow becomes supersonic $v(r)>c_s(r)$ somewhere.
\\
{\bf 4)}
For any radius $r_{\rm w}>0$
of the wire  there are always bound states
for some values of $m$ and $\omega$.

These results can be shown in analogy to the sound-ray case.
Let us first note that the Bernoulli theorem
($\vau^2/2+h(\varrho)=\rm const$ for free stationary flow)
implies that the enthalpy $h(\varrho)$ drops towards the center,
because $\vau$ increases.
Thus the pressure $p$ and the density $\varrho$ must decrease
near the wire.
Furthermore, for most fluids the Gr\"uneisen parameter
$\propto dc_{\rm s}/d\varrho$ is positive (e.g., for $^4$He II at $T=0$,
we have $dc_{\rm s}/d\varrho\in[2.2,2.9]\times c_{\rm s}/\varrho$,
\cite{Donnelly_data}), and therefore the speed of sound is also a
monotonically decreasing function of $r$.
Thus the term $\omega^2(1-1/c^2)$ in $\cal V$ is negative.
For sufficiently large $\omega$, one could get bound states for
$m=0$ via total reflection, see point {\bf 2}.
However, if the change in $c$ is small and thus the required frequencies
are too large, the underlying fluid dynamic description might not be
applicable anymore.

In contrast, the bound states for $m\neq0$ mentioned in point {\bf 4}
can occur for smaller values of $\omega$ and at larger length scales.
Let us consider the operator $\mathcal H$, where in the potential
$\mathcal V$ we may vary the parameters $m$ and $\omega$.
The remaining freedom of choosing $p_z$ can be used to adjust $\lambda$.
For an arbitrary test function $\psi$, the kinetic part
$\{\psi|{\cal D}|\psi\}>0$ of the expectation value of $\cal H$
is independent of $m$ and $\omega$, but for counter-rotating modes
the expectation value of $\cal V$ can be made arbitrarily negative.
For instance for $\omega=-m$ the expectation value of $\cal V$
is negative for all test functions and scales as $m^2$.
Thus, if $m^2$ is large enough, we get $\{\psi|{\cal H}|\psi\}<0$,
i.e., bound states must exist.

\paragraph*{Case of constant $\rho$ and $c$}

The arguments stated above prove the existence of the bound states for
large enough $m$, but they do not provide information about how large
$m$ must be for a given set-up and how $\omega$ depends on $p_z$,
for example.
To get this information, we numerically solve Eq.~(\ref{eigenvalue})
for the special case of constant $\rho$ and $c$.
For moderate velocities $\vau$ (sufficiently below the speed of sound),
this should be a reasonably good approximation.

\begin{figure}[h!]
\centering
\includegraphics[scale=0.37]{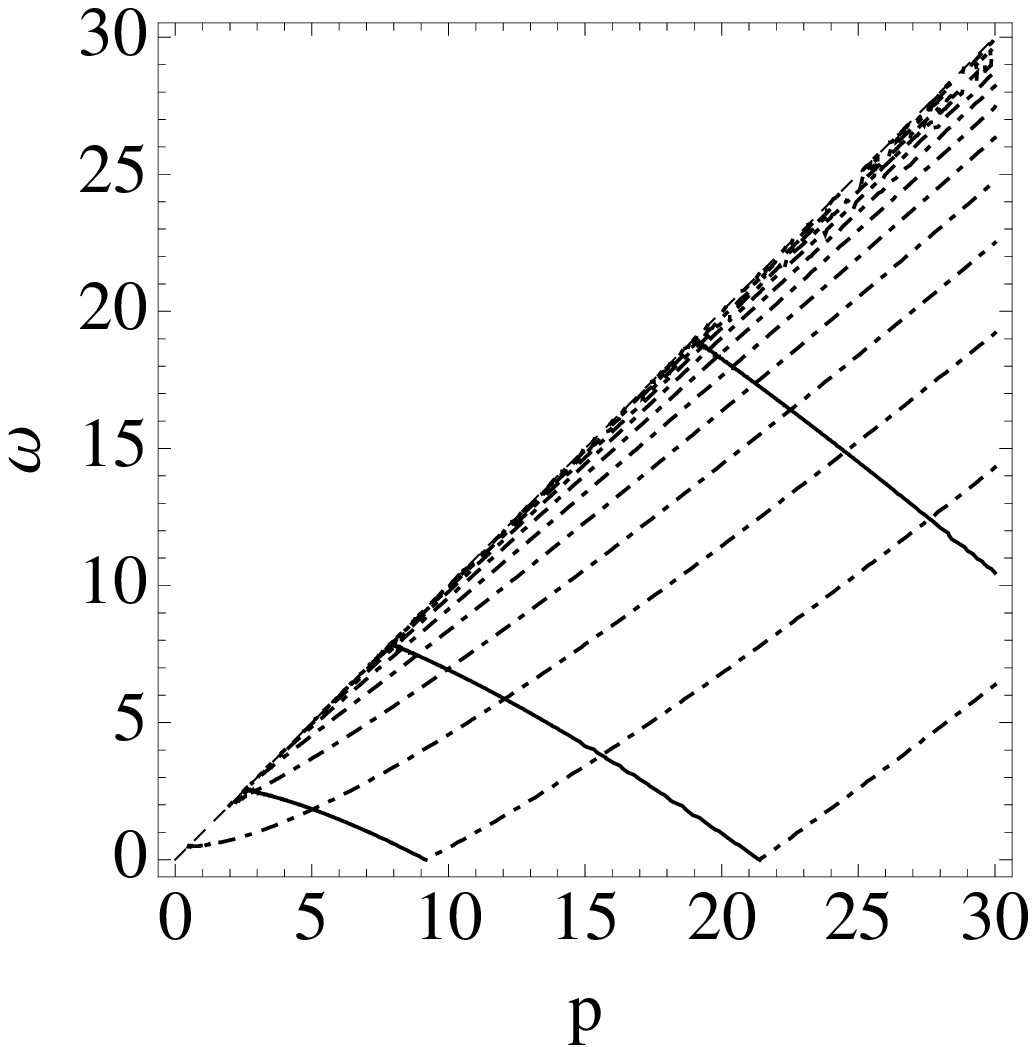}
\hfill
\includegraphics[scale=0.41]{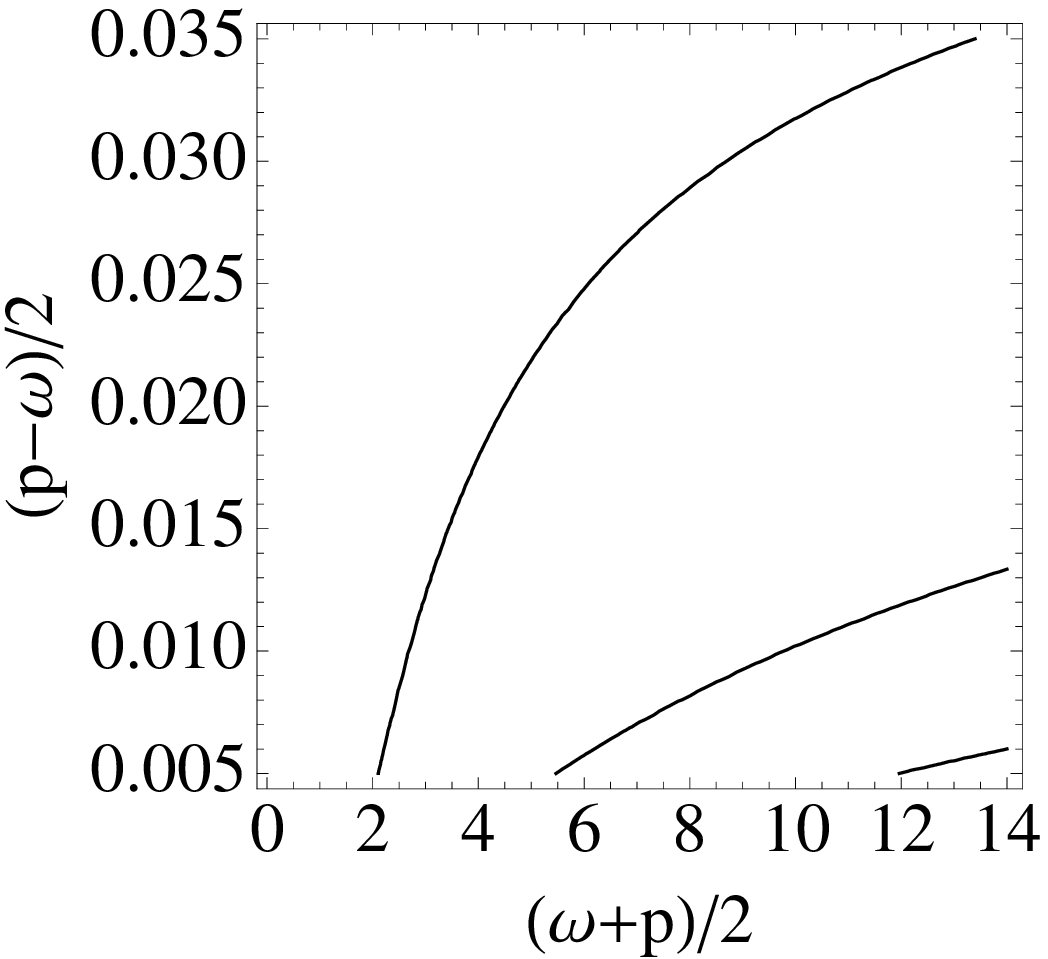}
\caption{Dispersion relations for bound states.
(Left) with ergoregion, $r_{\rm w}=0.3$; (solid) co-rotating waves $m=5$; 
(dashed) counter-rotating waves $m=-5$.
(Right) without ergoregion, $r_w=5$, rotated and zoomed view for
counter-rotating waves, $m=-3$. The region $\w\geqslant |p|$ 
is filled with scattering states.}
\label{disp}
\end{figure}

In Fig.~\ref{disp}, we plot examples of the dispersion relations
$\omega(p_z)$ of the bound states.
In the right plot, we choose $r_{\rm w}=5$, in which case that the
flow velocity does not exceed $c_\infty/5$ and thus the density should 
be constant up to a few percent.
In agreement with statements {\bf 2} and {\bf 3}, we only find
counter-rotating bound states, here we plotted modes with $m=-3$. 
In agreement with statement {\bf 1}, there is a gap for such modes, i.e.,
a minimum frequency, of $\omega_{\rm min}\approx 1.7$.
Above this gap, the dispersion relation is almost linear and very close 
to the $\omega=p_z$ line
 -- note the different scales on the $(p-\omega)/2$ and the 
$(p+\omega)/2$ axes -- 
which means that these modes propagate with almost the sound speed along 
the vortex.

For comparison, we plotted an example with an ergoregion $r_{\rm w}=0.3$
in Fig.~\ref{disp} (left).
Even though such a profile is probably hard to realize experimetally
(especially with $\varrho\approx\rm const$), one could imagine keeping
the density nearly constant by applying a suitable external potential
$V$ such that the stationary Bernoulli equation reads
$\vau^2/2+h(\varrho)+V=\rm const$.
Furthermore, one could rotate the cylinder with the same velocity as 
the innermost layer of the fluid in order to stabilize the set-up better.
%
%
This profile allows co-rotating bound states (here $m=5$)
which have a negative group velocity 
in addition to the counter-rotating modes (here $m=-5$).
In this case,
neither co- nor counter-rotating modes have a gap -- 
at discrete values of $p_z$, the frequency 
and thus the energy $E=\omega(\Phi|\Phi)$
vanishes.
Such zero-energy modes are not uncommon in problems involving 
perturbations of vortices \cite{Fabre} 
and may indicate an instability. 

\paragraph*{Conclusions}

For a vortex pinned at a wire, we studied sound propagation via Unruh's
analogy to gravity.
On general grounds, we predict the existence of bound states --
whispering gallery like modes -- based on the geometric acoustics
approximation as well as the full wave equation.

It should be possible to verify the characteristics of the bound states 
in experiments. 
%
For example, let us consider superfluid $^4$He with a typical speed
of sound $c_{\rm s}=2.4\times 10^4 {\rm cm}/{\rm s}$.
For a singly quantized vortex we get
$\kappa=1.6\times 10^{-4}{\rm cm^2}/{\rm s}$.
This leads to the length scale
${\mathfrak L}=6\times 10^{-9}{\rm cm}$
which is smaller than the van der Waals radius of Helium,
and to the frequency scale
${c_\infty^2}/{\kappa}=3.6\times 10^{12}{\rm Hz}$
above the roton frequency.
At these scales, fluid dynamics 
(which is the basis of our sound description) 
is not valid anymore.
Thus, for observing these bound states, it is probably better 
to trap many circulation around the wire,
which linearly increases $\kappa$ and thus ${\mathfrak L}$, 
while decreasing the frequency of the modes.

We remark that the family of bound states presented here is distinct 
from the phenomenon of Kelvin waves, known from normal- and superfluid 
dynamics. 
The vortex considered in this Letter is pinned and not allowed 
to move or be deformed. 
Accordingly, the dispersion relation (e.g., in  Fig.~\ref{disp}) 
of the modes considered here is much stiffer and more sound-like than 
that of the soft Kelvin waves with 
$\w(p_z)\sim\kappa p_z^2\log(p_z)/2$ for small $p_z$, 
see \cite{Donnelly_book}.
In fact, in the absence of ergoregion 
the bound states discussed here 
do not even exist for 
small $p_z$, where the above Kelvin formula is intended to be applicable.
%

\paragraph*{Acknowledgement}

Fruitful discussions with P.~Stamp, G.~Volovik, and W.G.~Unruh
are gratefully acknowledged.
This work was supported by the DFG (MA4851/1-1, SCHU 1557/1-3, SFB-TR12).


\end{document}